\documentclass{moriond}

\usepackage{tcolorbox}
\usepackage{verbatim}
\usepackage{amssymb} 
\usepackage[normalem]{ulem}

\def\Journal#1#2#3#4{{#1} {\bf #2}, #3 (#4)}

\newcommand{\HH}{{\cal H}}

\newcommand{\OmGW}{\Omega_{\rm GW}}

\newcommand{\ee}{\BoldVec{e}{}}
\newcommand{\dd}{{\rm \, d}}

\newcommand{\Eq}[1]{Eq.~(\ref{#1})}

\newcommand{\Fig}[1]{Fig.~\ref{#1}}

\newcommand{\FFig}[1]{Figure~\ref{#1}}
\newcommand{\Sec}[1]{Sec.~\ref{#1}}

\newcommand{\vw}{v_{\rm{w}}}
\newcommand{\cs}{c_{\rm{s}}}

\def\PLB{{\em Phys. Lett.}  B}

\def\PRD{{\em Phys. Rev.} D}

\newcommand{\is}[1]{\textcolor{green}{#1}}
\usepackage{soul} 
\newcommand{\sis}[1]{\textcolor{green}{\sout{#1}}}
\newcommand{\cis}[1]{\textcolor{green}{(#1)}}

\def\mco{\multicolumn}

\def\ra{\rightarrow}

\def\ko{K^0}

\def\be{\begin{equation}}
\def\ee{\end{equation}}
\def\bea{\begin{eqnarray}}
\def\eea{\end{eqnarray}}

\newcommand{\Photo}{\includegraphics[height=35mm]{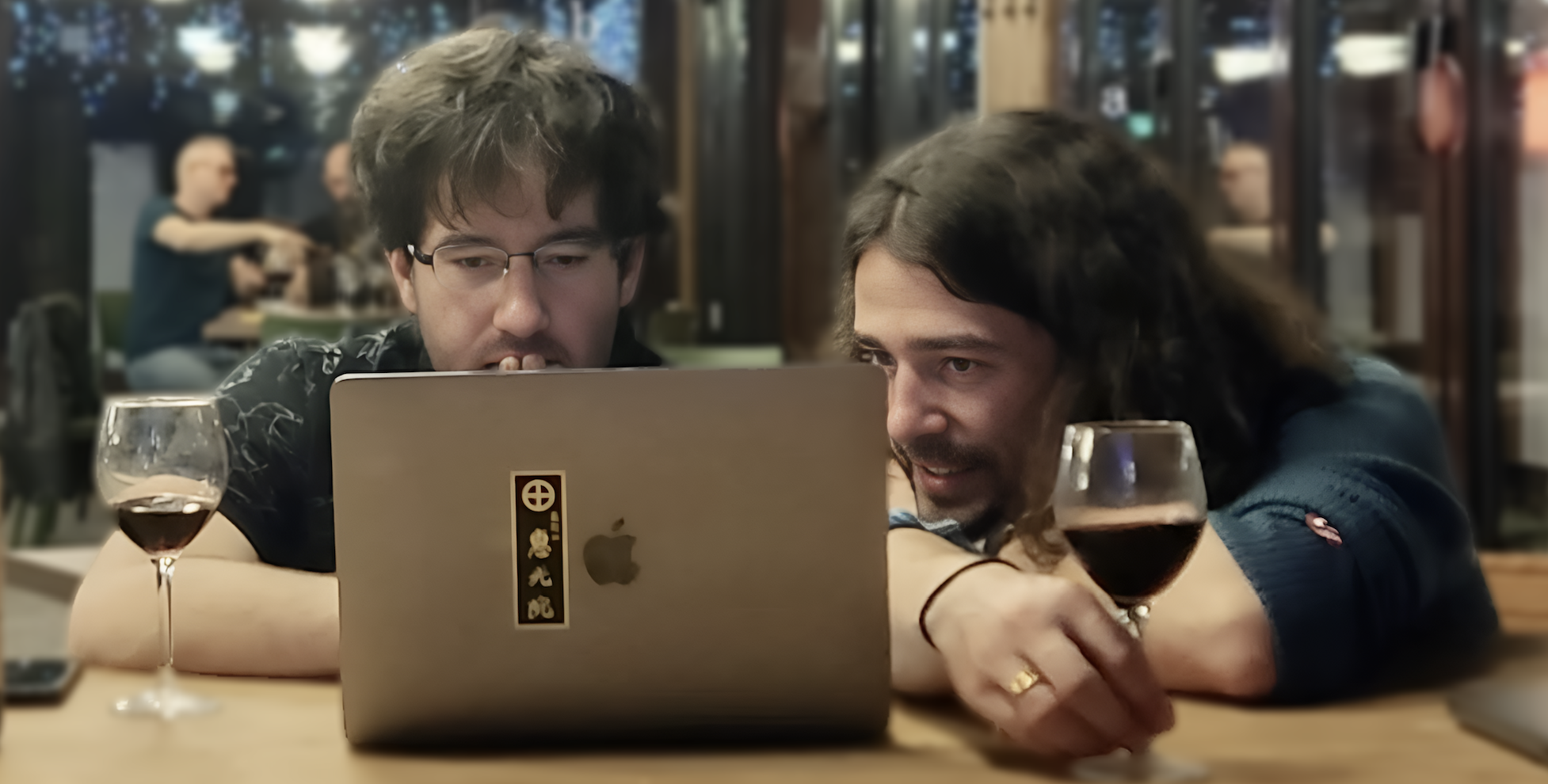}}

\begin{document}
\title{Gravitational wave spectra for cosmological phase transitions with non-linear decay of the fluid motion}

\author{Isak Stomberg\,$^a$ and Alberto Roper Pol\,$^b$ }
\address{$^a$ IFIC, Universitat de València-CSIC, C/ Catedràtico José Beltràn 2, E-46980, Paterna, Spain \newline
$^b$ D\'epartement de Physique Th\'eorique, Universit\'e de Gen\`eve,
CH-1211 Gen\`eve, Switzerland \newline}

\maketitle\abstracts{
We summarize the theoretical framework of gravitational wave (GW) production by bulk fluid motion induced by expanding broken-phase bubbles during a first-order phase transition.
Using a locally stationary unequal-time correlator (UETC) to model the decay of the source due to non-linearities, we provide templates for the resulting GW background that have been validated against data from Higgsless simulations.\,\cite{Caprini:2024gyk}
This UETC generalizes the stationary one considered in the sound-shell model—appropriate for linear sound waves whose kinetic-energy decay is negligible—to encompass the non-linear evolution of the compressional fluid motion beyond the sound-wave regime.
We demonstrate the implementation of templates based on this theoretical description
and the results from the Higgsless simulations\,\cite{Caprini:2024gyk} in the public Python package {\sc CosmoGW}, facilitating their use in experimental forecasts and parameter-estimation studies. The GW spectrum is delivered as a function of the key phase transition parameters: the wall velocity $v_w$, the strength $\alpha$, the nucleation rate $\beta$, and the source duration $\delta\eta_{\rm fin}$. \\
{\bf \em Contribution to the Gravitation session of the 59th Rencontres de Moriond.}
}

\section{Introduction}

The upcoming space‑based Laser Interferometer Space Antenna\,\cite{LISA:2017pwj} (LISA) will explore mHz frequencies of the gravitational wave (GW) spectrum and could be sensitive to cosmological backgrounds sourced by first‑order phase transitions (PTs) around the electroweak scale.
Since the LISA band will be populated by multiple astrophysical sources, both resolvable and
unresolvable,
in the form of a stochastic background, as well as other potential sources to the cosmological GW
background,\,\cite{Colpi:2024xhw}
reliable spectral templates are therefore essential. Plagued, however, by model uncertainties, methodological choices, and limited computational resources, current GW predictions from PTs often yield disparate results, complicating detectability forecasts for LISA and hindering parameter-inference studies.
Building on the Higgsless simulations\,\cite{Caprini:2024gyk,Blasi:2023rqi,Jinno:2022mie} and
a theoretical framework validated
by these simulations,\,\cite{Caprini:2024gyk} we provide an analytic parameterization that captures the time‑decaying nature of the GW source.
We present the implementation of these results in the public Python package 
\href{https://github.com/cosmoGW/cosmoGW}{{\sc CosmoGW}} and provide a \href{https://github.com/CosmoGW/CosmoGW/blob/main/tutorials/GWs_sound-waves.ipynb}{tutorial}
showing its use, comparing our results to previous templates for sound-wave production of GWs
in the literature.

\section{Gravitational wave production from decaying compressional sources}\label{sec:model}

The GW production from non-linear compressional fluid motion induced
by broken-phase bubbles in a first-order phase transition seems to be
well described by the assumption of a locally stationary unequal-time correlator (UETC) of the
source, according to the model and simulations of the PT dynamics following the Higgsless approach
presented in a recent work by the present authors and 
collaborators,\,\cite{Caprini:2024gyk} denoted as HL25 from now on.
In the following,
we review their model and simulation results, and provide templates
of the resulting GW background. 
Note that the assumption of a locally stationary UETC reduces to the stationary
one assumed in the sound-shell model when the kinetic energy
does not decay in time, i.e., in the linear regime (sound waves).\,\cite{Hindmarsh:2016lnk,Hindmarsh:2019phv,RoperPol:2023dzg}

In an expanding Universe, the GW spectrum produced by a source active during the radiation-dominated era and
characterized
by a locally stationary UETC
is given by the following expression\,\cite{Caprini:2024gyk}
\begin{equation}
    \OmGW (k) = \frac{1}{\rho_{\mathrm{tot}}} \frac{d \rho_{\mathrm{GW}}}{d \ln k} =3 \, F_{\rm GW}^0 \, \tilde \Omega_{\rm GW} \, K_{\rm int, exp}^2 \, \HH_\ast {\cal R}_\ast \,
    S(k {\cal R}_\ast)\,, \label{OmegaGW}
\end{equation}
where  $\rho_{\rm tot }$ is the total energy density of the Universe at present time,
$\rho_{\rm GW}$ is the GW energy density, $\HH_\ast$ is the conformal Hubble rate at the time of production,
$k$ is the comoving wave number,
$
    F_{\rm GW}^0 = \left(a_\ast/a_0\right)^{4} \left(H_\ast/H_0\right)^2
    \simeq 1.6 \times 10^{-5} \left(g_\ast/100\right)^{-1/3}
$ is the redshift factor, ${\cal R}_\ast = (a_\ast/a_0) R_\ast$
is the characteristic comoving length scale of the fluid with $R_\ast \beta \equiv (8 \pi)^{1/3} \max(\vw, \cs)$,
and $\tilde \Omega_{\rm GW} \sim {\cal O} (10^{-2})$ is an efficiency factor that has been
studied numerically~\cite{Caprini:2024gyk,Hindmarsh:2015qta,Hindmarsh:2017gnf} and
analytically~\cite{Hindmarsh:2016lnk,Hindmarsh:2019phv} for stationary sound waves.
$S(k)$ corresponds to the spectral shape, normalized such that $\int S(k) \dd \ln k = 1$.

To model the decay of compressional motion due to non-linearities in the system, especially when the PT
is strong, the UETC of the anisotropic stresses sourcing GWs,
under the locally stationary assumption, can be expressed at conformal times $\eta_1$ and $\eta_2$, as
\begin{equation}
E_\Pi (\eta_1, \eta_2, k) = 2 \, k^2 \, K^2 (\eta_+) \, f(\eta_-, k)\,, \label{eq:Epilocstat}
\end{equation}
where $\eta_+ = \frac{1}{2} (\eta_1 + \eta_2)$, $\eta_- = \eta_1 - \eta_2$, and $K=\rho_{\rm{kin}}/\bar \rho$ is the kinetic energy fraction.
Conformal time is denoted by $\eta$, such that $\dd \eta = a \dd t$, where $a$ is the scale
factor and $t$ is cosmic time.
The function $f(\eta_-, k)$ characterizes the  stationary UETC,
already considered in the sound-shell model used to study sound waves.\,\cite{Hindmarsh:2016lnk,Hindmarsh:2019phv,RoperPol:2023dzg}
The addition of a time-dependent $K^2 (\eta_+)$ in the UETC leads to the dependence of the GW amplitude on $K_{\rm int, exp}^2$ in \Eq{OmegaGW}, which
is~\cite{Caprini:2024gyk}
\begin{equation}
    K_{\rm int, exp}^2 = \HH_\ast^{-1} \int_{\eta_\ast}^{\eta_{\rm fin}}  \frac{K^2 (\eta_+)}{\eta_+^2}
    \dd \eta_+ \,. \label{K_decay}
\end{equation}
The source has been assumed to be present for a finite duration $\delta \eta_{\rm fin} = \eta_{\rm fin} - \eta_\ast \gg {\cal R}_\ast$ within the radiation-dominated era.
This is a reasonable assumption for kinetic energy densities $K \ll 1$
as the source is expected to last at least until
the shock formation time, $\delta \eta_{\rm fin} \gtrsim \eta_{\rm shock} \equiv {\cal R}_\ast/\sqrt{v_f}$,
where $v_f$ is the enthalpy-weighted root mean square velocity.\,\cite{Caprini:2024hue}
At the end of the PT, we assume an equation of state $p = \rho/3$ for the fluid,
such that the equations of motion become conformally flat and the numerical results of HL25
in Minkowski
space-time can be applied to an expanding Universe.\,\cite{Brandenburg:1996fc,RoperPol:2025lgc}
In their numerical simulations, it is found that the time evolution of the kinetic energy fraction
can be described as a decaying power-law $K(t) = K_0\,(t/t_0)^{-b}$ after the end of the
PT, with $b \geq 0$ indicating the decay rate and $t_0$ the time at which the PT completes in the simulations.\,\cite{Caprini:2024gyk}
For this power-law decay, \Eq{K_decay} can be expressed as 
\begin{equation}
    K_{\rm int, exp} = K_0^2 \Upsilon_b (\HH_\ast \delta \eta_{\rm fin})\,,
\end{equation}
where $\Upsilon_b$ is computed from the integral of the power law over time\,\cite{Caprini:2024gyk}
and reduces to the suppression factor $\Upsilon (x) = x/(1 + x)$ when there is no decay of the source, i.e., $b = 0$.\,\cite{RoperPol:2023dzg,Guo:2020grp}

Numerical simulations show that the spectral shape $S(k)$ can be parameterized as a doubly broken 
power law,\,\cite{Caprini:2024gyk} also used in previous templates for sound-wave production of GWs,\,\cite{Hindmarsh:2019phv,Caprini:2024hue,RoperPol:2023bqa}
\begin{equation}
\label{eq:shape function}
S(k,\,k_1,\,k_2)=S_0\, \left(\frac{k}{k_1}\right)^{n_1}\left[1+\left(\frac{k}
{k_1}\right)^{a_1}
\right]^{\frac{-n_1+n_2}{a_1}}\left[1+\left(\frac{k}{k_2}\right)^{a_2}\right]^{\frac{-n_2+n_3}{a_2}},
\end{equation} 
where $S_0$ is a normalization factor such that $\int S(k) \dd \ln k = 1$,
$n_1 = 3$, $n_2 = 1$, and $n_3 \gtrsim -3$ indicate the slopes at small, intermediate, and large frequencies
respectively, $k_1$ and $k_2$ indicate the scales where the spectral slopes change, and
$a_1 = 3.6$ and $a_2 = 2.4$ determine the smoothness of the transition between slopes found in the simulations of HL25.
The value of $k_1 {\cal R}_\ast$ is found to have a
small dependence on the PT parameters, while $k_2$ is found to depend on the sound-shell thickness $\Delta_w = \xi_{\rm shell}/\max(v_w, c_{\rm s})$ for weak PTs ($\alpha = 0.0046$), as described in previous 
work.\,\cite{Hindmarsh:2019phv,Hindmarsh:2015qta,Hindmarsh:2017gnf,Caprini:2024hue,Hindmarsh:2013xza,Caprini:2019egz}
This dependence is not apparent for intermediate and strong PTs.
On average, the numerical values found in HL25 are $k_1 {\cal R}_\ast \simeq 0.4 \times (2 \pi)$ and $k_2 {\cal R}_\ast \simeq \pi/\Delta_w$, $k_2 {\cal R}_\ast \simeq 2 \pi$, and $k_2 {\cal R}_\ast \simeq \pi$ for
weak, intermediate, and strong PTs, respectively.

Following this model, validated by the simulations of HL25, we can
generate the GW spectrum predicted for any first-order PT described by the set of parameters $\{v_w, \alpha, \beta/H_\ast, \delta \eta_{\rm fin}\}$ once that the efficiency $\tilde \Omega_{\rm GW}$, the kinetic
energy fraction $K_0$, and the decay rate $b$ are known.
These quantities have been computed numerically in HL25 and are provided within \href{https://github.com/cosmoGW/cosmoGW}{{\sc CosmoGW}}, which allows to
extrapolate the numerical results to different values of $v_w$ and $\alpha$.
Note that $\beta/H_\ast$ determines the fluid length scale $R_\ast$ and the function $\Upsilon_b$.\,\cite{Caprini:2024gyk}

\section{Numerical results}\label{sec:results}

Simulations of the dynamics of a first-order
phase transition with an exponential
probability rate of nucleation $P(t) \propto e^{\beta t}$ using the Higgsless approach have recently been conducted for wall velocities $v_w \in [0.32, 0.8]$ and strengths $\alpha\in\{0.0046,\, 0.05,\, 0.5\}$.\,\cite{Caprini:2024gyk}
These simulations capture the GW production from the full dynamics of the fluid motion, potentially
including non-linear effects such as turbulence (vortical and acoustic) and shock formation.
The template that is developed based on the locally stationary
UETC, summarized in \Sec{sec:model}, is validated with the simulations in the time interval spanning
from $t_{\rm init} - t_0 \approx 5/\beta$ until $t_{\rm end} - t_0 \approx 21/\beta$ in Minkowski space-time, which
are used to determine the GW efficiency, $\tilde \Omega_{\rm GW}$, the
fraction of kinetic energy density, $K_0$, and the power-law decay
in time, $b$. 
These results mark an important generalization beyond the sound-shell model by capturing the effect of non-linear dynamics on the GW amplitude,
allowing to accurately predict, for the first time, the GW amplitude
for strong PTs ($\alpha = 0.5$).
These simulations have clearly demonstrated that capturing the full non-linear dynamics is crucial for accurate GW predictions, hereby adequately reflected and contained in the presented templates.

\begin{figure*}[t]
    \centering
    \includegraphics[width=0.99\columnwidth]
    {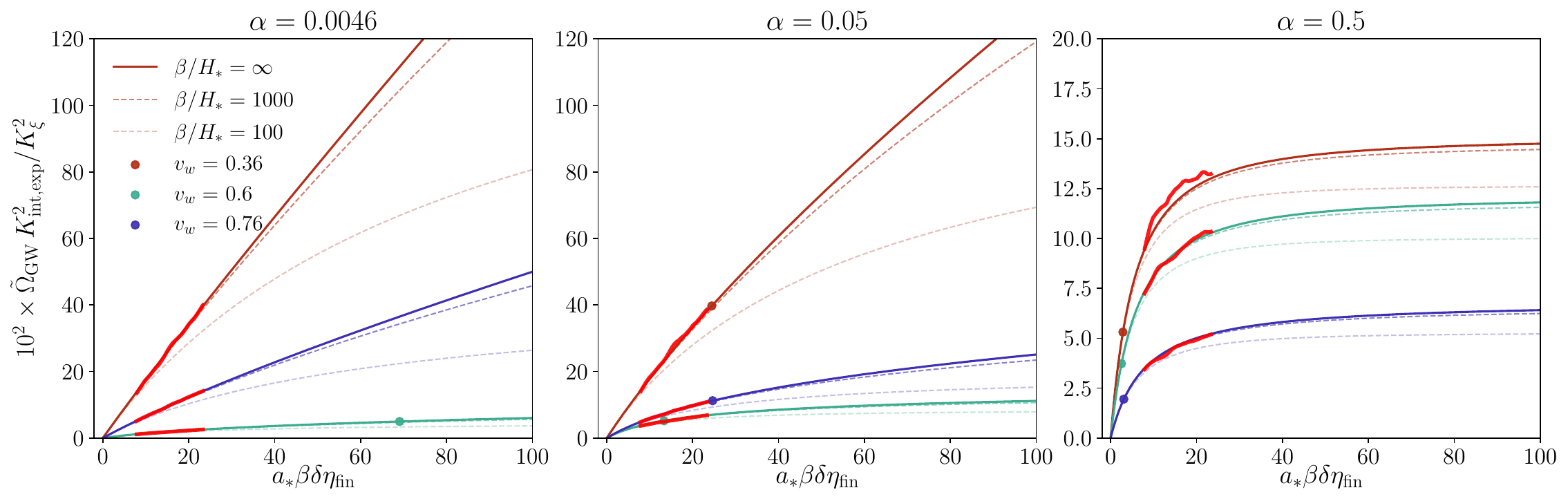}
    \caption{
    \label{fig:GWmodel}
    Dependence of the GW efficiency $\tilde \Omega_{\rm GW}$ with the conformal source duration
    $a_\ast \beta \delta \eta_{\rm fin} 
    = a_\ast \beta \eta_{\rm fin} 
    - a_\ast \beta \eta_\ast = a_\ast \beta \eta_{\rm fin}
    - \beta/H_\ast$ found in HL25.
    $\tilde \Omega_{\rm GW}$ is shown normalized by the reference value\, 
    $\tilde \Omega_{\rm GW} \simeq 10^{-2}$ and by $K_\xi^2 R_\ast \beta$,
    $K_\xi$ being the kinetic energy fraction of single bubbles.
    This figure corresponds to Fig.~8 of HL25.} 
\end{figure*}

Figure~\ref{fig:GWmodel} shows the dependence of the GW ampltitude with the source duration
$\delta \eta_{\rm fin}$ in Minkowski space-time ($\beta/H_\ast \to \infty$) following the locally
stationary UETC, compared to the
numerical results of HL25, extrapolating
the results to times before $t_{\rm init}$ and beyond $t_{\rm end}$, assuming that
the GW production starts at $t_\ast = t_0 < t_{\rm init}$.
Using the model presented in \Sec{sec:model} and the conformal invariance of the fluid equations,\,\cite{Brandenburg:1996fc,RoperPol:2025lgc} we apply the numerical results to estimate the
dependence for different values of $\beta/H_\ast = \{100, 1000\}$ in an expanding Universe.

A few notes are in order:
\begin{enumerate}

\item In weak ($\alpha = 0.0046$) and some intermediate ($\alpha = 0.05$) PTs,
the damping found in the simulations is small, and it could be
affected by numerical accuracy.\,\cite{Caprini:2024gyk}
However, we expect that for values of $\beta/H_\ast \lesssim 100$ (note from \Eq{OmegaGW} that GW amplitude
is suppressed for larger $\beta/H_\ast$), the saturation amplitude of the GW spectrum
is governed by the Hubble expansion, $\Upsilon (\delta \eta_{\rm fin} \gg 1) \to 1$, thus reducing the dependence on the source duration.

\item In strong PTs, the large decay rate ($b > 1$) found in the simulations leads to a saturated GW amplitude in a time scale similar to the simulation duration even in Minkowski space-time.
This is the most phenomenologically relevant case for LISA, as large $\alpha$ is required for
detectability (see \Fig{fig:GWtemplates}).
Therefore, even if the locally stationary UETC model 
would not describe the fluid dynamics at later times
due to, for example, the dominance of vortical turbulence,\,\cite{Kosowsky:2001xp,Caprini:2009yp,Niksa:2018ofa,RoperPol:2019wvy,RoperPol:2022iel,Auclair:2022jod}, the estimated GW amplitude taking $\delta \eta_{\rm fin} \to \infty$ is only modified
by a 10--20\% at most.
Including expansion makes the saturation amplitude to be reached at even earlier times.

\item In previous analyses of sound-wave production of GWs, the source is assumed to be active only until the shock formation time.
Since the decay of the source is captured in the presented model,
which accurately describes the GW amplitude found
in the simulations of HL25,
our estimates can be used beyond the shock formation time.
In particular, for strong PTs ($\alpha = 0.5$), the power-law decay is observed until the end of the simulations, which end at
around 5 shock formation times (see \Fig{fig:GWmodel}).

\end{enumerate}

\section{Template implementation and the {\sc CosmoGW} code}\label{sec:cosmogw}

\begin{figure*}[t]
    \centering
    \includegraphics[width=.85\columnwidth]{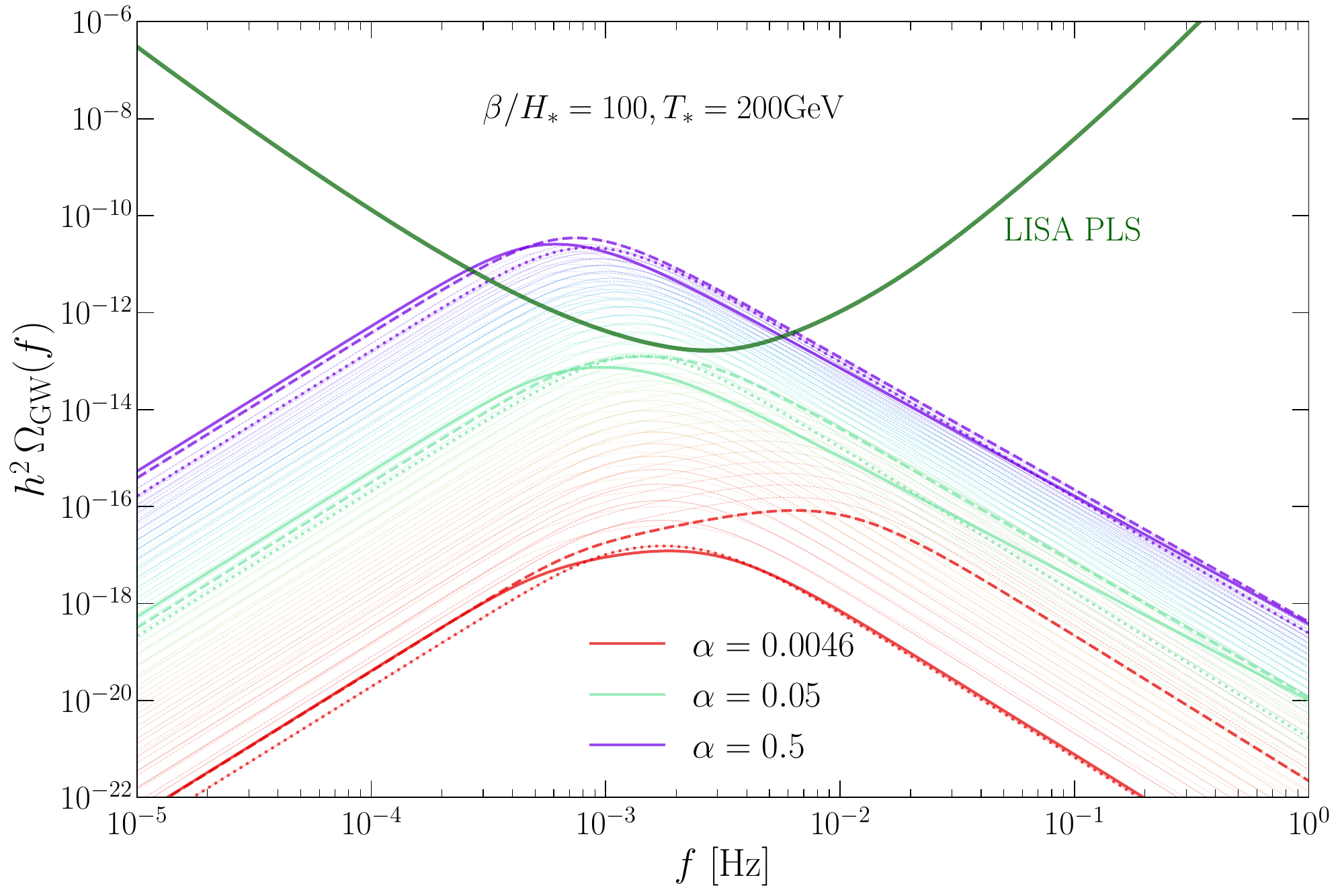}
    \caption{
    \label{fig:GWtemplates}
    Present-time GW spectrum produced by a first-order phase transition at the electroweak scale ($T_\ast \simeq 200$ GeV) computed using the locally stationary UETC
    presented in \Sec{sec:model} and the numerical results of HL25 for $\alpha = \{0.0046, 0.05, 0.5\}$ (thick lines), interpolated to intermediate values of $\alpha$ (thin lines), for $\beta/H_\ast = 100$, and for $v_w = \{0.36, 0.6, 0.76\}$.
    The GW spectra are compared to the power-law sensitivity (PLS) of LISA with an SNR of 10
    and for 4 years of observation.
    A \href{https://github.com/CosmoGW/CosmoGW/blob/main/tutorials/GWs_sound-waves.ipynb}{tutorial} with different templates has been made available as part of the 
    \href{https://github.com/cosmoGW/cosmoGW}{{\sc CosmoGW}} code.
    } 
\end{figure*}

We have implemented the template based on the locally stationary UETC model of Eq. (\ref{OmegaGW})
in the public \href{https://github.com/cosmoGW/cosmoGW}{\sc CosmoGW}
Python package.\,\cite{cosmogw} The implementation takes as inputs
$\{v_w,\, \alpha,\,\beta/H_\ast, N_{\rm shock}\}$, such that the source duration is $\delta \eta_{\rm fin} = N_{\rm shock} R_\ast/v_f$ and interpolates
between the parameters $\{\tilde \Omega_{\rm GW},\, k_1,\, k_2,\, n_3,\, K_0,\, b\}$
extracted from the numerical simulations of HL25
with $v_w\in [0.32, 0.8]$ and $\alpha\in\{0.0046,\, 0.05,\, 0.5\}$.\,\footnote{As additional simulation results become available, the corresponding support points will be added for more reliable interpolation.} 
A minimal example demonstrating its usage is shown in the code snippet below, accompanied by
a \href{https://github.com/CosmoGW/CosmoGW/blob/main/tutorials/GWs_sound-waves.ipynb}{tutorial}, where previous templates in the literature
are also described.
\begin{tcolorbox}[
  colback=white, 
  colframe=black, 
  boxrule=0.5pt, 
  sharp corners,
  left=1mm, 
  right=1mm, 
  top=0.5mm, 
  bottom=0.5mm,
  fontupper=\tiny\ttfamily
]
\begin{verbatim}
!pip install cosmoGW
from cosmoGW import *
import matplotlib.pyplot as plt; import numpy as np; import astropy.units as u

# Define phase transtions parameters
alphas = np.array([0.5, 0.05, 0.0046])
vws    = np.array([0.36, 0.56, 0.8])
betas  = np.array([100])
T = 200*u.GeV # electroweak temperature scale
g = 100       # relativistic degrees of freedom at the electroweak scale

# Range of frequencies normalized with the fluid length scale s = fR*
s = np.logspace(-2, 2, 1000)

# Compute the GW spectrum
freqs, OmGWs = GW_templates.OmGW_spec_sw(s, alphas, betas, vws=vws, expansion=True, Nsh=100,
                                         model_efficiency="higgsless", model_K0="higgsless",
                                         model_decay="decay", interpolate_HL_decay=True, model_shape="sw_HLnew",
                                         interpolate_HL_shape=True,
                                         interpolate_HL_n3=True, redshift=True, T=T, gstar=g)
                                        
# Plot the GW spectrum
cdict = {0.0046: "steelblue", 0.05: "red", 0.5: "orange"}
plt.figure(figsize=(8, 8))
for i,alpha in enumerate(alphas):
    for j,vw in enumerate(vws):
        plt.loglog(freqs[:, j], OmGWs[:, j, i], color=cdict[alpha], lw=1.5, alpha=vw, label=f"$v_w = {vw}$, $\\alpha = {alpha}$")
        
# Compare to LISA’s power law sensitivity PLS with SNR = 10 and 4 years of observation
f_LISA, _, LISA_OmPLS = interferometry.read_sens(SNR=10, T=4)
plt.plot(f_LISA, LISA_OmPLS, lw=3, alpha=.5, color="darkgreen")
plt.xlabel(r"$f$ [Hz]", fontsize=24)
plt.ylabel(r"$h^2 \, \Omega_{\rm GW} (f)$", fontsize=24)
plt.legend(fontsize=12, frameon=False)
\end{verbatim}
\end{tcolorbox}
\FFig{fig:GWtemplates} shows the resulting GW spectra found using the model of \Sec{sec:model}
and interpolating the results from
the Higgsless simulations presented in HL25.
The GW amplitude is estimated
using the decay rate $b$ of HL25 and taking the long-duration limit $\delta \eta_{\rm fin} \HH_\ast \gg 1$ (as the decay is being modeled).
In previous templates, e.g., the one used by the LISA
CosWG,\,\cite{Caprini:2024hue}
the duration of the stationary source (i.e., with no decay) is taken as $\delta \eta_{\rm fin} \HH_\ast \simeq \min(1, R_\ast/v_f)$.
Another difference with previous templates is that
the second peak in HL25 is found to be almost independent on the
wall velocity
for $\alpha = 0.05$ and $0.5$, potentially due to the development
of non-linearities.

\section{Conclusions}\label{sec:conc}

We have presented a practical and physically grounded gravitational-wave (GW) template derived from the locally stationary unequal-time correlator (UETC) framework introduced and validated against data from the Higgsless simulations of HL25.\,\cite{Caprini:2024gyk} This template captures the key features of GW production from first-order cosmological phase transitions (PTs), particularly the non-linear decay of 
the fluid compressional motion beyond the sound-wave regime.
Utilizing data extracted from HL25 across a wide range of PT parameters, we have implemented the resulting template in the open-source \href{https://github.com/cosmoGW/cosmoGW}{\sc CosmoGW} Python package,\,\cite{cosmogw} enabling fast and reliable predictions of the GW spectrum capturing the full non-linear dynamics as a function of the wall velocity $v_w$, the PT strength $\alpha$, the rate of nucleation $\beta/H_*$, and the source duration $\delta \eta_{\rm fin}$. We observed that, for strong PTs, when the decay is strong, the dependence on the source duration is reduced as the GW
amplitude is close to saturation by the end of the
numerical simulations of HL25. The numerical implementation thus provides the community with consistent and user-friendly access to the latest GW predictions from first-order PTs based on state-of-the-art methods.

Our work is ongoing, and we plan to update the {\sc CosmoGW} package to reflect the state-of-the-art additional support points in PT parameter space and incorporating refined predictions and models as they become available. Previous models and templates are also available for comparison.
Future efforts will focus on extending the parameter-space coverage, refining the modeling of the spectral shape in expanding backgrounds, and validating the template over even longer source durations.

\section*{Acknowledgements}
We extend our appreciation to the 59th Rencontres de Moriond organizing committee for their continuous efforts to deliver the most inspiring conference, and to our collaborators Chiara Caprini, Ryusuke Jinno, Thomas Konstandin, and Henrique Rubira. IS acknowledges support by the Generalitat Valenciana through the Programa Prometeo for Excellence Groups, grant CIPROM/2022/69 ``Sabor y origen de la materia.'' ARP acknowledges support by the Swiss National Science Foundation
(SNSF Ambizione grant \href{https://data.snf.ch/grants/grant/208807}{182044}).

\bibliography{refs}        

\end{document}